\documentclass{article}[]
\usepackage{epstopdf}
\usepackage{graphicx,tikz,endnotes,amsmath,bbm}
\usepackage[natbibapa]{apacite}
\usepackage{multirow}
\usepackage[flushleft]{threeparttable}
\usepackage{float}
\usepackage{longtable,siunitx}
\usepackage[colorlinks]{hyperref}
\usepackage{etoolbox}
\usepackage[utf8]{inputenc}

\usepackage{xcolor,graphicx,float,amsmath,tikz,etoolbox,environ,pgfplots,pgfplotstable,eurosym,afterpage,pdfpages}

\usetikzlibrary{calc,positioning,shadows.blur,decorations.pathreplacing,matrix,shapes.misc,arrows,dateplot,fit,arrows, decorations.markings}
\usepackage{subcaption}

\usepackage{array}
\usepackage{makecell}

\usepackage{hyperref}
\hypersetup{
    colorlinks=true,
    linkcolor=blue,
    citecolor=blue,
    filecolor=blue,
    urlcolor=blue,
}

\title{Systematically Monitoring Social Media:\\ The case of the German federal election 2017}

\author{Sebastian Stier, Arnim Bleier, Malte Bonart,\\ Fabian Mörsheim, Mahdi Bohlouli, Margarita Nizhegorodov, \\ Lisa Posch, Jürgen Maier, Tobias Rothmund, Steffen Staab
}

\date{\normalsize \textbf{Keywords:} election campaigning; Facebook; Twitter; political communication; social media; Bundestag\\[0.35cm] 
Published in \emph{GESIS Papers} $2018\mid04$\\
 \url{http://nbn-resolving.de/urn:nbn:de:0168-ssoar-56149-4}}

\newcommand{\tab}{\hspace{4mm}} 
\newcommand{\ttab}{\hspace{2mm}}

\pgfplotsset{compat=newest}

\tikzset{
        brace/.style = { decorate, decoration={brace, amplitude=5pt} },
       mbrace/.style = { decorate, decoration={brace, amplitude=5pt, mirror} },
        label/.style = { black, midway, scale=0.5, align=center },
     toplabel/.style = { label, above=.5em, anchor=south },
    leftlabel/.style = { label,rotate=-10,left=.5em,anchor=north }, 
   rightlabel/.style = { label,left=-1.0em,anchor=west,align=right}, 
        round/.style = { rounded corners=2mm },
}

\tikzstyle{vecArrow} = [thick, decoration={markings,mark=at position
   1 with {\arrow[semithick]{open triangle 60}}},
   double distance=1.4pt, shorten >= 5.5pt,
   preaction = {decorate},
   postaction = {draw,line width=1.4pt, white,shorten >= 4.5pt}]
\tikzstyle{innerWhite} = [semithick, white,line width=1.4pt, shorten >= 4.5pt]

\begin{document}

\maketitle



\section{Introduction}

The campaign for the German parliament (\emph{Bundestag}) 2017 was finally the one in which party strategists and observers regarded social media not just as experimental and in essence peripheral venues, but rather as important arenas where elections are won or lost. The year before, Donald Trump won the presidency in the U.S., which was attributed to his authentic Twitter use and a skillful mobilization of supporters via social media. 
Right-wing populist forces on the rise in Germany like the AfD or Pegida similarly use social media to bypass media gatekeepers and reach sympathetic target audiences \citep{Stier.2017a}. So-called ``fake news'', social bots (semi-automated accounts) and online propaganda (e.g., orchestrated by Russia), accompanied by a growing mistrust of legacy media in the wake of the refugee crisis threatened to impact the campaign. The agenda-setting power of online media (\citealp{Neuman.2014}) became apparent when an online campaign led by party activists helped to propel Martin Schulz and the SPD to a parity with the CDU in public opinion polls in March 2017 (``Schulzzug''). In the campaigning arena, parties applied innovations like micro-targeting at a larger scale in order to harness the persuasive potential of social media. 
Importantly, the intense -- maybe even overproportional -- coverage of the aforementioned phenomena by the mass media contributed to the perception of an increased political role of social media.

In order to understand these processes and also to keep economic actors like Facebook and Twitter accountable for their political influence, it is essential for acade\-mia to systematically explore new digital data sources. Yet, online social networks are complex and intransparent sociotechnical web environments \citep{strohmaier2014}. Therefore, it is a considerable task to collect digital trace data at a large scale and at the same time adhere to established academic standards. In the context of political communication, important challenges are (1) defining the social media accounts and posts relevant to the campaign (\emph{content validity}), (2) operationalizing the venues where relevant social media activity takes place (\emph{construct validity}), (3) capturing all of the relevant social media activity (\emph{reliability}), and (4) sharing as much data as possible for reuse and replication (\emph{objectivity}).  

The present collaborative project by GESIS -- Leibniz Institute for the Social Sciences and the E-Democracy Program of the University of Koblenz-Landau conducted such an effort. We concentrated on the two social media networks of most political relevance, Facebook and Twitter. These platforms have different architectures, user bases and usage conventions that need to be taken into account conceptually and methodologically. Section \ref{sec:related_work} discusses previous work related to our endeavor. In Section \ref{sec:polcomm}, we lay out what kinds of activities we define as part of the ``political communication space'' on Facebook and Twitter. Section \ref{sec:data_coll} outlines our data collection. In Section \ref{sec:analysis}, we present exploratory findings of how political communication on the Bundestag campaign unfolded on social media. Section \ref{sec:datasharing} discusses how we share the main output of the project, the ``BTW17 dataset'' and lists of election candidates and their social media accounts. We conclude with a critical reflection of our efforts and an outline of future research avenues in Section \ref{sec:conclusion}.

\section{Related Work}
\label{sec:related_work}

A burgeoning literature has analyzed the role of social media during election campaigns. In this section, we first structure related work according to the logic of data collection. Then we summarize methodological research that has revealed several pitfalls when collecting social media data. Finally, we discuss related work on the Bundestag campaign 2017. The predominant focus in the existing literature clearly lies on Twitter use during election campaigns.\footnote{It is not our goal to review this literature in its entirety. Instead, we focus on commonalities in the predominant data collection strategies. For a comprehensive literature review on Twitter use during election campaigns, see \citet{Jungherr.2016a}.} Generally, one can distinguish between audience-centered and elite-centered data collections. 

Audience-centered designs capture tweets containing a set of specific hash\-tags (``\#btw17'') or keywords related to a campaign (``cdu'', ``merkel''). The goal is to reveal how interested users participate in political debates. From this body of research, we have a pretty good understanding of the dynamics that unfold during election campaigns in the public Twitterverse. A particular focus lies on ``second screening'' during TV debates \citep{Trilling.2015,Freelon.2015,Lin.2014,Vaccari.2015}, influential users \citep{Freelon.2015,Juergens.2011,Dubois.2014}, attempts to predict election results \citep{Tumasjan.2010,DiGrazia.2013} and the identification of central political topics \citep{Bruns.2011,Trilling.2015,Jungherr.2016bb}.

Elite-centered designs concentrate on a well-defined subset of the Twitterverse, namely specific accounts of interest. In contrast to audience-centered designs focusing on the demand side of politics, elite-centered studies concentrate on the supply side of politics, i.e., politicians, parties or journalists. Moreover, not only the tweets of these users mentioning a political keyword are of interest, but rather their overall behavioral patterns on social media. Here, studies focused on the adoption of platforms by politicians \citep{Quinlan.2017,Vergeer.2013}, the (partisan) structure of online networks of candidates \citep{Lietz.2014,Aragon.2013} and their resonance with audiences \citep{Kovic.2017,Yang.2017,Nielsen.2013}.

Studies of Facebook are more sparse, which is likely due to technical and privacy constraints. Most Facebook profiles and communication are private, whereas on Twitter, most posts and user profiles are public. Since retrieving posts via keyword search is only possible for (the few) public posts, studies of election campaigns on Facebook necessarily have to concentrate on users with public pages. Thus, most electoral studies are elite-centered in that they concentrate on profiles of politicians and parties (e.g., \citealp{Levon.2016,Caton.2015,Williams.2013,Kovic.2017,Nielsen.2013}). From this preselection, the analysis sometimes still moves to the audience, for example the politically active users in the comments sections of these pages (e.g., \citealp{Freelon.2017}).


Our project also draws on previous studies that critically assessed the data provided by Twitter's Application Programming Interface (API). \citet{Driscoll.2014} performed a comparison between the publicly accessible Streaming API (providing up to 1\% of live Twitter traffic) and the Gnip PowerTrack Firehose dataset provided by Twitter (granting full access to the Twitter data stream). They found that during periods of public contention like election campaigns or protests, the Streaming API returns biased data because the rate limit tends to be surpassed when a social phenomenon generates a lot of public interest. Similarly, \citet{Morstatter.2013} found that as a researcher increases the parameters (keywords) of interest, the coverage of the Streaming API decreases. In the most systematic evaluation to date, \citet{Tromble.2017} compared the Streaming API and the Search API to a Firehose dataset as a ground truth. While they identified serious biases in the Search API, they found that the Streaming API returns ``nearly all'' relevant tweets when rate limits are not reached. Based on this methodological research, we design a data collection scheme that is robust to the biases inherent to the Twitter API (see Section \ref{sec:data_coll}).

In contrast to Twitter, the Facebook Graph API allows the collection of all non-deleted posts for an unlimited research period. Thus, researchers can design lists of accounts they want to monitor and even collect data ex post, after an election campaign has unfolded. This has been the standard approach in academic works, which, however, misses activities such as posts, comments and likes that got deleted in the meantime. In a first systematic analysis of deletion rates, \citet{Bachl.2017} showed that approximately 18\% of user generated content on German political Facebook pages had been deleted over the span of eight months. Among posts by political actors themselves, only 2.3\% of posts could not be retrieved anymore.\footnote{We could not incorporate these insights in our project, as Bachl presented first findings in a conference presentation only on 22 September 2017.} Our own data collection from Facebook generated two independent datasets that are susceptible to these limitations to various degrees (see Section \ref{sec:data_coll}).

As for studies related to the current Bundestag campaign, \citet{Schmidt.2017} collected social media accounts of all candidates. In his study, he describes the adoption of social media by candidates and their follower relationships on Twitter. 
Yet, in addition to this metadata that can also be accessed ex post, our project also collected the contents of candidates' tweets as well as candidates' interactions with other users in real time. Another project collected data at a larger scale, albeit exclusively on Twitter \citep{Kratzke.2017}.\footnote{We are aware that various other teams collected specific Twitter data for concrete research questions as well.} Apart from being limited to one platform, this project only collected data for 360 politicians, mostly sitting members of the Bundestag. Such a convenience sample only allows for limited substantive analyses. 

In the following section, we describe which target concepts we consider essential to be covered during an election campaign. Furthermore, we describe how they can be operationalized on social media. In our definition of the ``political communication space'', we chose a hybrid approach that integrates both, audience-centered and elite-centered research designs.

\section{Defining the Political 
Communication Space}
\label{sec:polcomm}

\begin{figure}[H]
\centering
\resizebox{.8\textwidth}{!}{
\begin{tikzpicture}

\draw (0.4,1.2) -- (0.4,-2.3);
\draw (4.1,1.2) -- (4.1,-2.3);

  \node at(2.25,1)      {\textbf{Target concepts}};
  \node at(6.00,1)       {\textbf{Twitter}}; 
  \node at(-1.7,1)      {\textbf{Facebook}};

\node (cCandidates) [draw, minimum width=3.0cm, dashed, align=left, anchor=north] at(2.2, 0.45) {\bf Candidates\\[0.5mm] \bf Organizations\\ \tab Parties\\ \tab Gatekeepers};

\node (cTopics) [below=of cCandidates, yshift=0.8cm,draw, minimum width=3.0cm, rounded corners=.55mm, align=left,dashed] {\bf Important topics};

\node (tTimelines) [draw, minimum width=3.0cm, rounded corners=.55mm, dashed, align=left, anchor=north] at(6.2, 0.45) {account timelines\\ retweets\\[0.75mm] @-mentions};

\node (tHashTags) [below=of tTimelines, yshift=0.75cm,draw, minimum width=3.0cm, rounded corners=.55mm, align=left,dashed] {hashtags\\ string matches};

\node (fPages) [draw, minimum width=3.0cm, rounded corners=.55mm,  align=left, anchor=north,dashed] at(-1.6, 0.45) {\bf pages\\ \ttab posts\\ \ttab comments \\ \ttab likes};

\draw [vecArrow] (3.72, -0.3) to (tTimelines);
\draw [vecArrow] (.72, -0.4) to (fPages);
\draw [vecArrow] (cTopics) to (4.72, -1.9);

\draw [mbrace] (7.8,-0.45) -- (7.8,0.4)
                 node[rightlabel] {\LARGE ID based};

\draw [mbrace] (7.8,-2.18) -- (7.8,-0.5)
                 node[rightlabel] {\LARGE string based};
\end{tikzpicture}}
\caption{The political communication space and its operationalization.}
\label{fig:collection}
\end{figure}
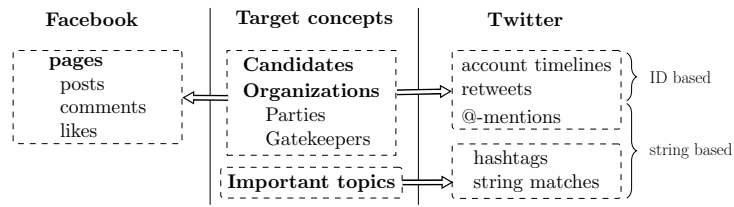

Our goal was to collect all publicly available political communication related to the Bundestagswahl on Facebook and Twitter. To this end, we define three target concepts on which political communication is typically centered: (1) politicians, here Facebook pages and Twitter accounts of candidates in the election campaign, (2) Facebook pages and Twitter accounts of political parties and gatekeepers such as media organizations, and (3) keywords denoting central political topics on Twitter. This holistic conceptualization not only covers the most important actors, but also aims to capture the online engagement of regular citizens with politics. For politicians and parties, we have confined our data to those parties that had, based on polls, a realistic chance to win seats in parliament.\footnote{Included parties are AfD, Bündnis 90/Die Grünen, CDU, CSU, FDP, Linke and SPD.} Figure \ref{fig:collection} visualizes the conceptualization which we will explain in detail in this section. 

\subsection{Candidates}
The official list of candidates eligible for the election was not published by the \emph{Bundeswahlleiter} until mid of August, just a month before election day and therefore right in the middle of the campaign. However, our goal was to cover campaign activity as early as possible. Hence, we had to devote considerable efforts to first compile lists of candidates. This process was guided by the specifics of the electoral system for the Bundestag, mixed-member proportional representation. 50 percent of members of parliament are elected via relative majority in electoral districts (direct candidates). The other 50 percent are chosen from party lists elected by the regional divisions of parties in the 16 federal states (list candidates).

In case of list candidates, the process of identification was helped by the fact that most parties at state level published this information on their websites rather early. For the direct candidates, a comprehensive (web) search was required, because the nomination process is based at the local level. The preparation of the candidate list was a continuous process with the benchmark being the official list published in August.\footnote{Our original search missed only two percent of candidates (52 of the 2,516 final candidates). On the other hand, we identified 45 individuals who did not end up being an official candidate. The latter resulted 
mostly from disorganized/fluid processes and conflicts within party organizations, especially in the case of the AfD.}

Based on the candidate lists we identified the respective Facebook pages and Twitter accounts of the candidates. The main source were the websites of politicians and parties linking to social media accounts. In other cases we used Facebook's and Twitter's search and look-up functionality along the following (soft) criteria: Do the name, self-description and/or photos correspond to the politician? Is the account mentioned/linked in relevant conversations? Is the content of post real (or satire, e.g., in the case of Angela Merkel parody accounts on Twitter)? For Twitter, we collected one account per person and for Facebook, we collected up to two accounts, regardless of being a page or profile. 

After the election, we compared our data to two similar datasets that became public at the time.\footnote{These are a collection made by the Tagesspiegel featuring candidates' Facebook and Twitter accounts (\url{http://wahl.tagesspiegel.de/2017/kandidatenbank}) and data from the Open Knowledge Foundation on Facebook accounts (\url{https://docs.google.com/spreadsheets/d/1l1JO13aDCcTFHoVD0KjWFHSCe6h3joPjBiSJxKcZFkE}).} While we covered a substantially larger number of accounts (199\% more than the Tagesspiegel, 195\% more than the Open Knowledge Foundation), we still added 151 Facebook accounts and 52 Twitter accounts from these datasets to our final list.

The data on candidates including available attributes and their social media accounts 
can be downloaded from \citet{DBK.2018}.

\subsection{Organizations: Political Parties and Gatekeepers}

Besides candidates, there are additional influential accounts on social media during election campaigns, most importantly, accounts by political parties and news media. Likewise, their Twitter accounts and Facebook pages were researched for the construction of an ``organizations'' dataset.

The list of relevant party actors contains all accounts of the national parties, their caucuses in the Bundestag (``Bundestagsfraktion''), the parties in the federal states (``Landesparteien'') and youth organizations (e.g., the Jusos). Parties at all these different levels predominantly focused on the Bundestag campaign during our research period. At the same time, future research should also incorporate the local party branches (e.g., CDU Cologne). 

In addition to these party actors, we also compiled lists of accounts belonging to the right-wing protest movement Pegida that generate a lot of engagement on German social media \citep{Stier.2017a}. It was to be expected that these accounts would form part of the right-wing media ecosystem during the election campaign.

In order to compile a list of media accounts, we crowdsourced the account names of German media present on Facebook and Twitter. On the crowdsourcing platform CrowdFlower\footnote{\url{www.crowdflower.com}}, we asked German crowdworkers to find links to German media accounts (``mainstream media" as well as ``alternative media") that report on political topics. The rationale behind using (and paying) crowdworkers for this task was to construct a diverse set of accounts that also captures the long tail of media accounts. In total, we collected 6,211 responses from crowdworkers on Twitter media accounts (2,815 unique accounts) and 4,774 responses on Facebook media accounts (2,049 unique accounts). All media accounts with more than one mention were then manually looked through by the first author who screened out non-German accounts and non-related genres like YouTube Stars. The final list of media pages on Facebook contains 285 accounts, the list of media accounts on Twitter comprises 310 accounts.

\subsection{Political Topics}
\label{sec:pol_topics}

Finally, to monitor central political topics we built a list of related keywords (``selectors''). We predefined a broad set of relevant keywords before the election and decided not to add additional selectors during the campaign. The main reason for this decision is that if a researcher includes new trends, she will always be too late as these can only be identified when they are already popular on social media or in public debates. Thus we first opted for the more systematic approach with a fixed set of topics, before retrospectively capturing additional campaign topics, as explained later.

The selection of keywords related to a topic is a process prone to human biases \citep{King.2017}. Thus, our choice of selectors is certainly not objective but rather reflects ex ante expectations by the project team which topics and actors would become important during the campaign. We also had to consider two potentially distorting aspects. First, some keywords were used very frequently in other languages and would have flooded our data collection. For instance, we found out that ``fdp'' is an abbreviation very frequently used in French and Portuguese tweets. Thus we added some keywords with a hashtag, e.g. ``\#fdp'' that most often refers to the German party. Second, some surnames are not unique identifiers for German politicians. The word ``gabriel'', e.g., is a word used very often in Spanish, thus we included the full name ``sigmar gabriel''.

We minimize the number of missing relevant messages as we have such a holistic conceptualization of political actors and topics. Many messages on newly emerging topics still contain mentions of known candidates, party names or leading candidates. However, the share of relevant communication neglected by our definition of the political communication space is impossible to evaluate -- a bias that is inherent to any such data collection effort.

As a starting point for the construction of our selectors list, we used the keywords already collected in the social media data collection conducted by GESIS in 2013 \citep{btw2013}. Many of these are universally relevant (e.g., ``finanzpolitik'' or ``tvduell''). Furthermore, collecting these again enables us to directly compare two federal election campaigns in Germany. 

Analogous to 2013, we added keywords referring to political institutions and democracy in general, i.e., the German \emph{polity} (Appendix \ref{appendix:selectors1_polity}). Based on the list from 2013 and an investigation of the recent political agenda, we added keywords on issues (\emph{policy}). This includes traditional fields such as social policy or economic policy, but also topics of particular importance in 2017. For the list of selectors related to policy, we refer to Appendix \ref{appendix:selectors2_policy}.

We also captured keywords related to \emph{politics}, in particular on the ongoing election campaign (Appendix \ref{appendix:selectors3_politics}). Among these are popular generic hashtags like ``btw17'' and political TV events (``tvduell'') that typically generate a lot of attention on Twitter \citep{Jungherr.2016bb,StierBleier.2017,Freelon.2015,Trilling.2015}. 

Last, our selectors list comprised keywords related to political parties (Appendix \ref{appendix:selectors4_parties}). This includes the names of the established parties that are not only @-mentioned as organizational accounts very frequently, but also mentioned in free text (as ``political topics'' in our logic) without an @-prefix. Moreover, we include the names of the leading candidates (``Spitzenkandidaten''). This is of particular importance as the leading candidate of the CDU Angela Merkel does not have a Twitter account. We also included the names of cabinet members in the federal government and the prime ministers of German states (``Bundesländer''), as of beginning of July 2017. 

During the campaign, we assembled keywords that were relevant but not captured by our original list.\footnote{These included ``weidel'', the AfD politician who only rose to prominence during the campaign, ``flüchtling'' and ``migranten'' which we missed in our original list, and the party campaign hashtags ``fedidwgugl'', ``traudichdeutschland'', ``holdirdeinlandzurück'', ``denkenwirneu'', ``lustauflinks'', ``darumgruen'', ``zeitfürmartin''.} In order to capture these ex post, we used a script to scrape tweets containing these keywords from the Twitter interface allowing us to track topics until their first emergence. Yet this approach has the limitation that retweets are excluded from the interface and that deleted tweets are not available anymore. We were able to capture up to 100 retweets of these original tweets by retrospectively connecting to the Twitter REST API. Through this process, we added additional 2,136,620 tweets to our data. While this approach is not as systematic as the live streaming and has several inherent biases, we opted for still adding these tweets to our collection. We will evaluate the performance of this procedure with data we are currently buying from the data vendor Twitter Enterprise.

\section{Data Collection}
\label{sec:data_coll}

At the outset, it is important to note that we only collect publicly available information from social media. We collect these digital traces of user activity from the Facebook Graph API and the Twitter Streaming API.
In principal, all online interactions of users such as politicians and politically active individuals can be relevant for social science research. Building on the above defined political communication space, we are first interested in the tweets and Facebook posts of political candidates and organizations. Second, we collect the engagement of users with these contents -- retweets and @-mentions on Twitter, comments, shares and likes on Facebook. Third, we likewise consider keywords in messages. Finally, we also collect data from Twitter on follower relationships. 
The following subsections elaborate on how we technically monitored this political communication space.

\subsection{Data Retrieval from Twitter}

While Twitter can provide a transparent source of data, achieving this transparency requires to be precise on the data collection process. 
Building on the set of account names as described above, we used Twitter's Streaming API\footnote{\url{http://developer.twitter.com/en/docs/tweets/filter-realtime}} to collect messages sent by candidates, replies and retweets of these messages as well as messages sent to candidates (@-mentions), starting on 5 July. Using the same method, we collected the analogous messages for political parties and gatekeepers (the selection logic ``organizations'' described before). Furthermore, we captured messages containing at least one keyword from our list of political topics. The employed software includes Tweepy\footnote{\url{http://tweepy.org}} and Twitter4J\footnote{\url{http://twitter4j.org}} for retrieval as well as MongoDB\footnote{\url{http://mongodb.com}} for storing the data. Overall, we took extraordinary precautions to ensure the completeness of the collected data such as independent parallel collections by the project teams at our two institutions. 

Furthermore, we retrieved the follower graph for the monitored candidate accounts on 27 June 2017, 14 September 2017, 1 November 2017 and 9 January 2018. That way we know who followed politicians and whom they followed at different time points during the campaign. 
We retrieved the much larger follower graph for party and organizations accounts on 1 July 2017 (collecting 48 million connections).

\subsection{Data Retrieval from Facebook}
As elaborated above, the data that can be accessed and monitored publicly differs between social media platforms. On Facebook, a researcher can only collect data from public pages. As a consequence, political posts from non-public accounts (that many candidates also used) and conversations among Facebook friends or in closed groups can not be analyzed. The public portion of Facebook data can be accessed via the Facebook Graph API.\footnote{\url{http://developers.facebook.com/docs/graph-api}}

Given these constraints, we define the political communication space on Facebook as posts, comments and likes on public political pages by candidates and organizations. We set up two data collections. The first scheme collected data using an eight-day rolling time window starting on 15 August. This means that comments and likes on a given post were collected eight days after its creation. The rationale behind this approach is that most activity happens in the few days after a post had been created. We also set up a second dataset by retrieving Facebook data from all target accounts after the campaign. However, we recently got aware of the findings of \citet{Bachl.2017} who showed that such ex post data collections suffer from considerable deletion rates. Thus, we opted to report data from the first, continuous collection made during the campaign -- even though its temporal coverage is more limited.

\section{Exploratory Analysis}
\label{sec:analysis}

This section presents several exploratory analyses of our dataset covering the research period from 6 July to 30 September, one week after the election. We first evaluate available language filters that should be applied when analyzing a national election campaign. Next, we explore our dataset, focusing on (1) temporal patterns, (2) attention patterns on Twitter, i.e., which topics were the most salient in public discussions on the campaign and (3) activities and engagement by political parties.

\subsection{Data Cleaning}
As discussed in Section \ref{sec:polcomm}, every conceptualization of the political communication space will undoubtedly include posts unrelated to German politics (e.g., ``özdemir'' is the name of a leading candidate from the German Green party, but also a popular name in Turkey). This ``noise'' is less of a problem on Facebook, since all posts in our dataset have been contributed to a Facebook page relevant to German politics. Even if a user comments in another language choosing an entirely non-political topic on a page of one of our target accounts, this user still has intentionally chosen to contribute to the German political communication space. 

However, on Twitter, where we take out tweets from a large universe of messages posted in many different languages, ambiguities in the meaning of keywords become more of a problem. Moreover, international debates on German leaders or the campaign are not relevant or even distorting in the context of most social science research questions that still focus on the national arena. Many tweets on Angela Merkel, for instance, are coming from outside of Germany. These are related to her role in world politics, her relationship with Donald Trump or are very critical of her refugee policies (those tweets are often coming from the U.S. alt right).
\begin{table}[H]
\centering
\caption{Evaluation of language filters}
\label{tab:language}
\begin{tabular}{lll}
& \textbf{False positives}               & \textbf{False negatives}               \\ \hline
Twitter language interface DE & 14 (0.028\%)    & 121 (0.242\%) \\
Twitter machine learning DE   & 2 (0.004\%)     & 15 (0.03\%)  
\end{tabular}
\end{table}

In order to filter out tweets from non-German users and also reduce the amount of topically non-related ``noise'', we applied and evaluated two filtering approaches.\footnote{One additional source of ``noise'' is non-human activity. Our approach filters out bot activity in languages other than German which targets trending hashtags with spam and advertisements. Moreover, a recent study found that the percentage of messages sent by political bots on Twitter on the German election was only minor \citep{Kollanyi2017junk}. Note, however, that this study was based on an ad hoc sample of one million tweets on only select hashtags. Additionally, the authors define ``highly automated accounts'' as users tweeting more than 50 times per day on the election, which is certainly debatable.} The first approach restricted the dataset to only those tweets by users who chose the interface language German in their Twitter settings \citep{Jungherr.2016bb}. Second, we selected only tweets that were labeled as German by Twitter's machine learning algorithm. Then we randomly sampled and coded 4$\times$500 tweets from our dataset in order to assess the false negative and false positive rates of each approach. Results are reported in Table \ref{tab:language}.

It becomes clear that the Twitter machine learning approach is superior to a filtering based on the interface language. A non-trivial percentage of users who are tweeting in German on the election campaign are running their interfaces in other languages -- most often English. Meanwhile, the machine learning misclassifies only 2 of 500 non-German tweets as German and 15 of 500 German tweets as non-German. In total, the language of only 0.017\% of tweets is classified incorrectly by Twitter's machine learning algorithm. Therefore, we use the tweets that have been labeled as German by Twitter for our exploratory analysis of the Bundestag election campaign.

\subsection{Temporal Patterns}
We start our presentation of results with a description of the Facebook and Twitter datasets. The first purpose is to describe the raw data we collected before moving to more substantive analyses.

\begin{figure}
	\centering
     \begin{subfigure}[b]{0.49\textwidth}
     \includegraphics[width=\textwidth]{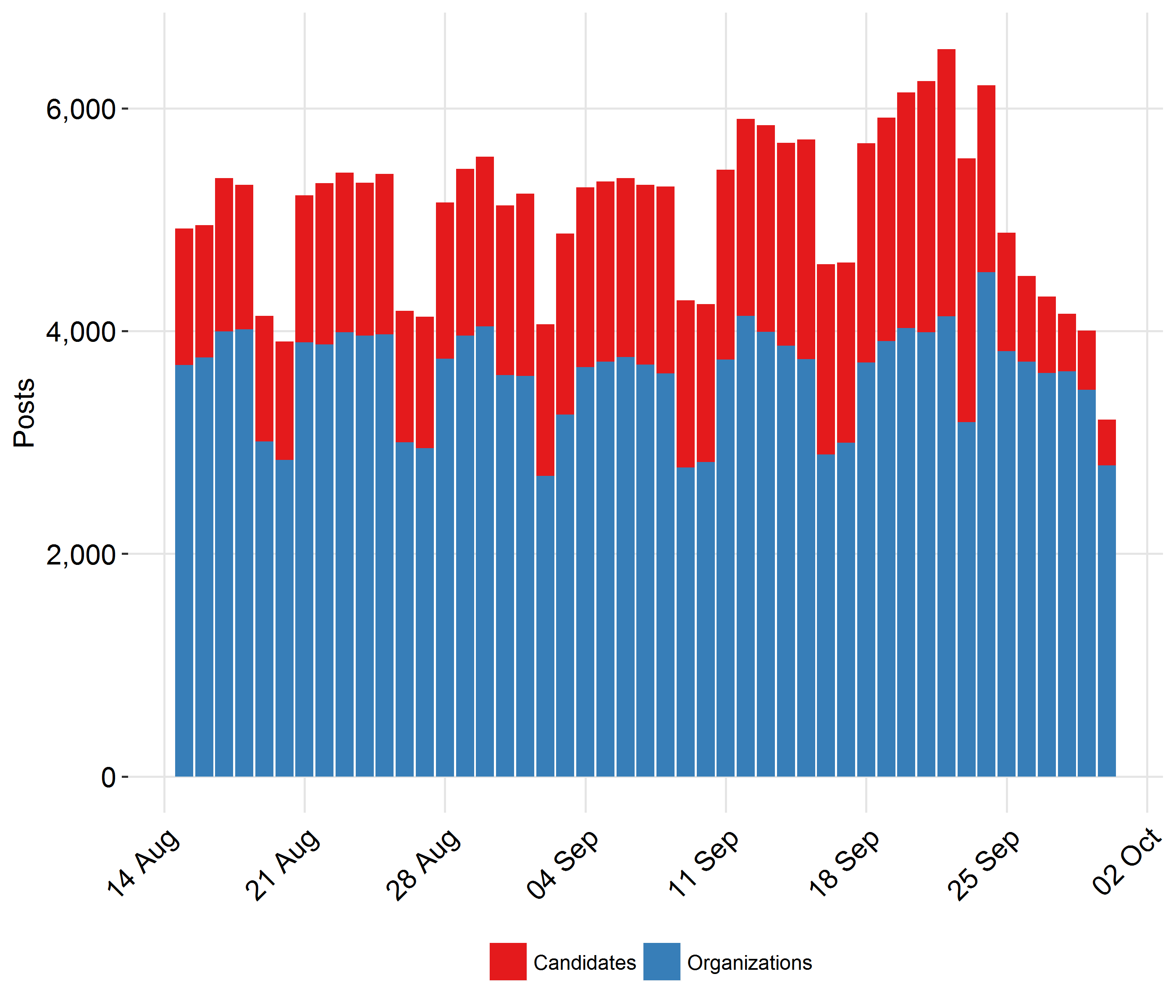}
        \caption{Posts}
    \end{subfigure}
    \begin{subfigure}[b]{0.49\textwidth}
        \includegraphics[width=\textwidth]{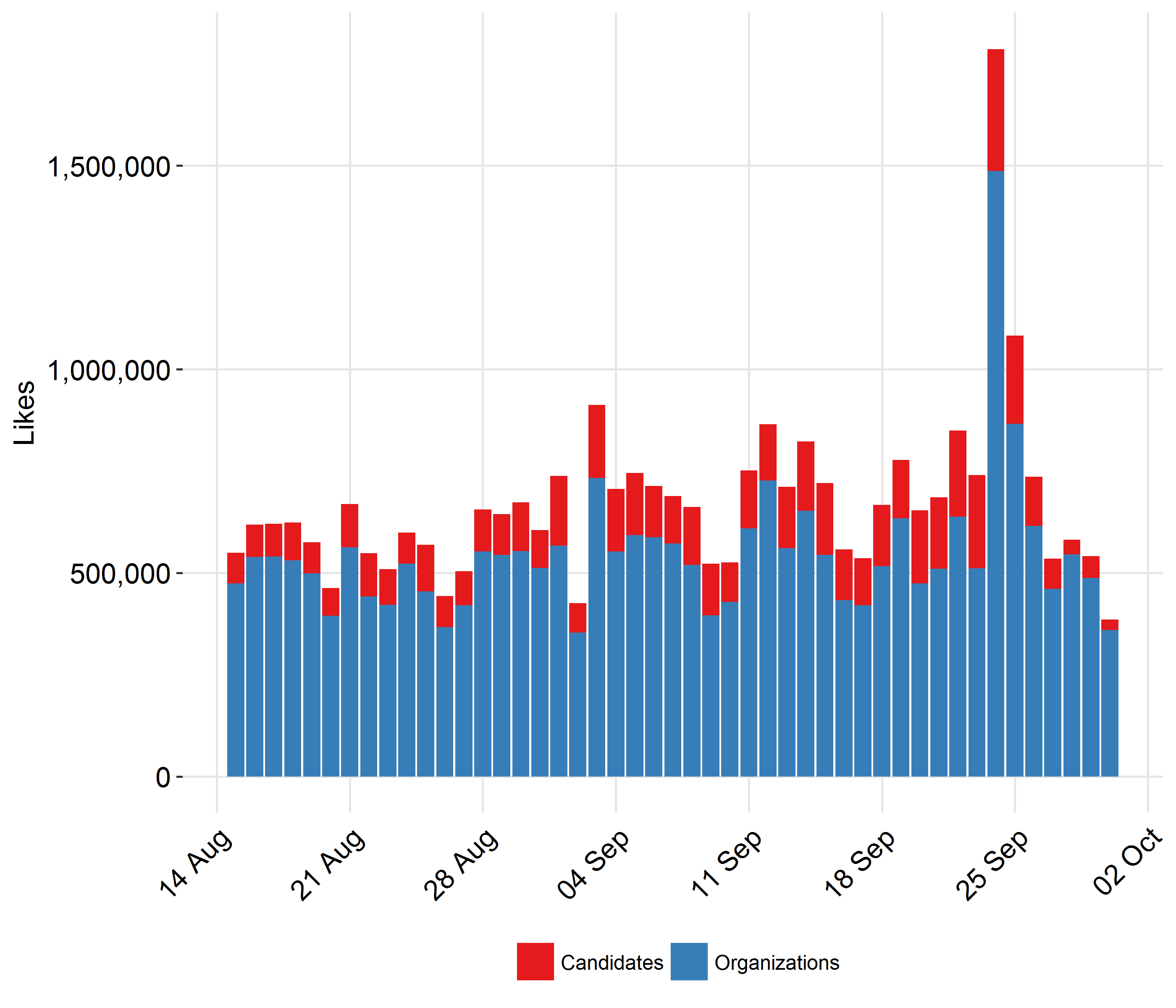}
       \caption{Likes}
    \end{subfigure}
     \begin{subfigure}[b]{0.49\textwidth}
        \includegraphics[width=\textwidth]{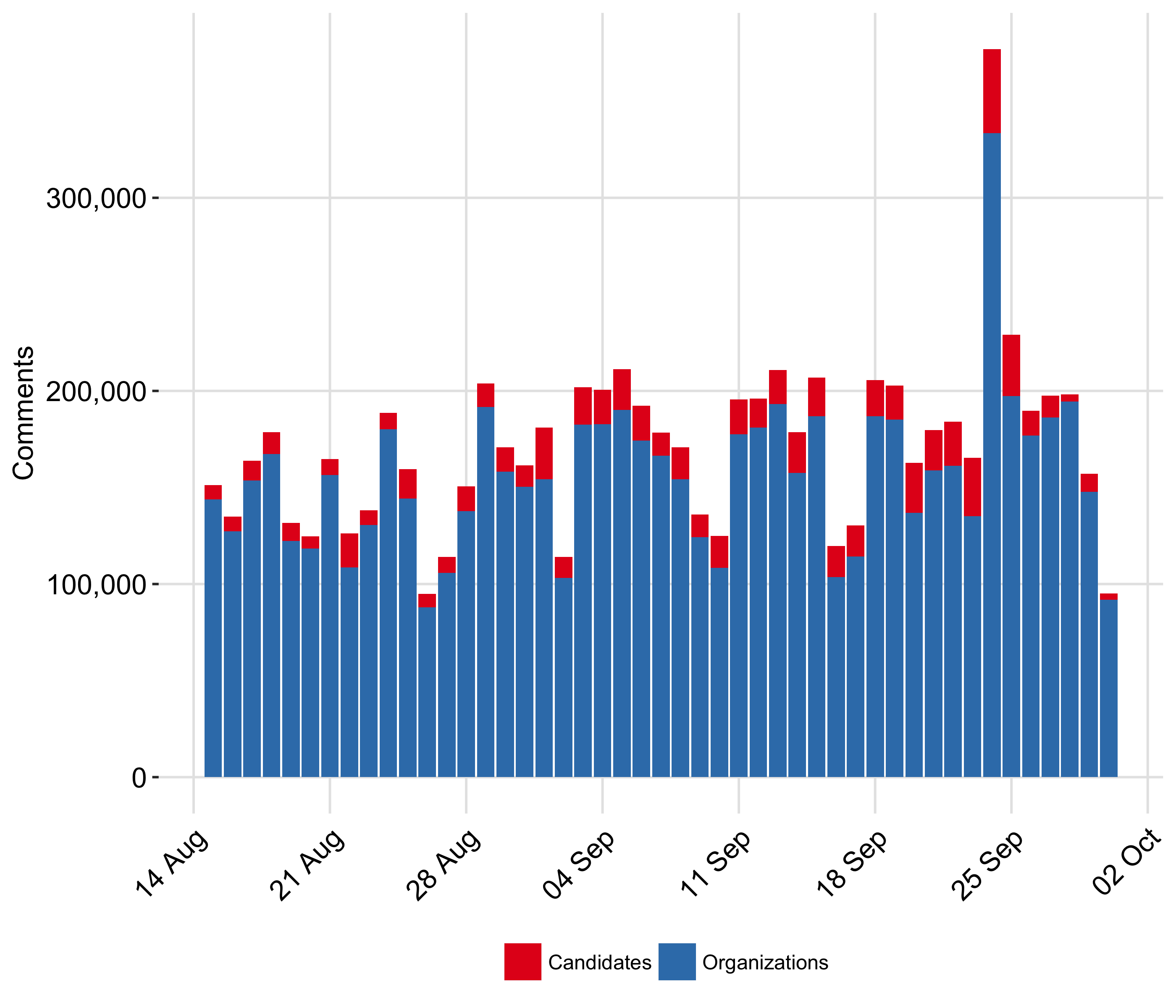}
        \caption{Comments}
     \end{subfigure}
   \caption{Facebook activities in our dataset by selection logic.}
	\label{fig:fb_daily}
\end{figure}

Figure \ref{fig:fb_daily} shows the daily Facebook activity over time as measured by posts, comments and likes. As is to be expected, the activity clearly increases with election day on 24 September approaching. On election night, Facebook posts by candidates, party organizations and gatekeepers received more than 1.5 million likes and more than 300,000 comments.

\begin{figure}
	\centering
    \begin{subfigure}[b]{0.49\textwidth}
        \includegraphics[width=\textwidth]{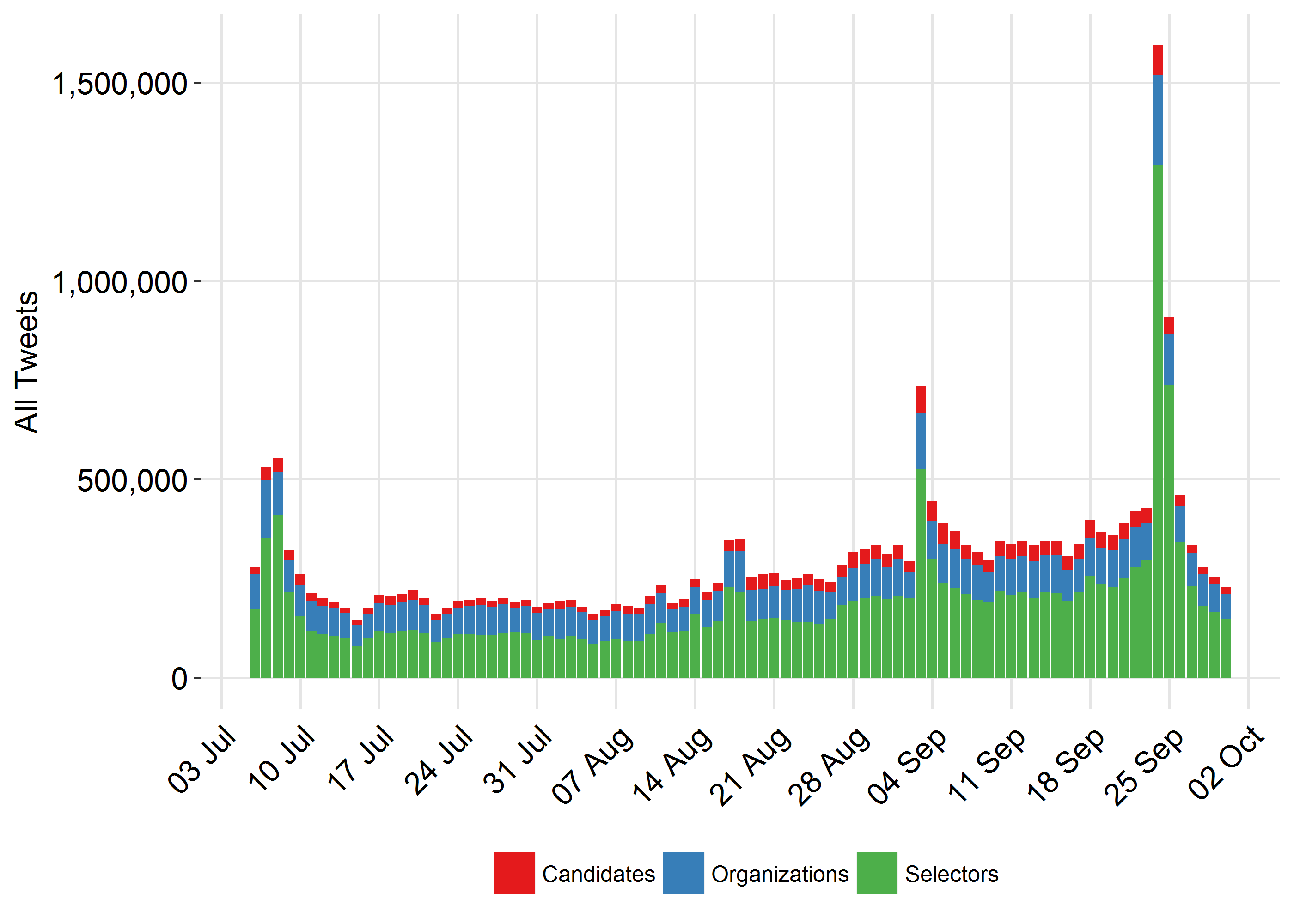}
        \caption{Unfiltered}
    \end{subfigure}
    \begin{subfigure}[b]{0.49\textwidth}
        \includegraphics[width=\textwidth]{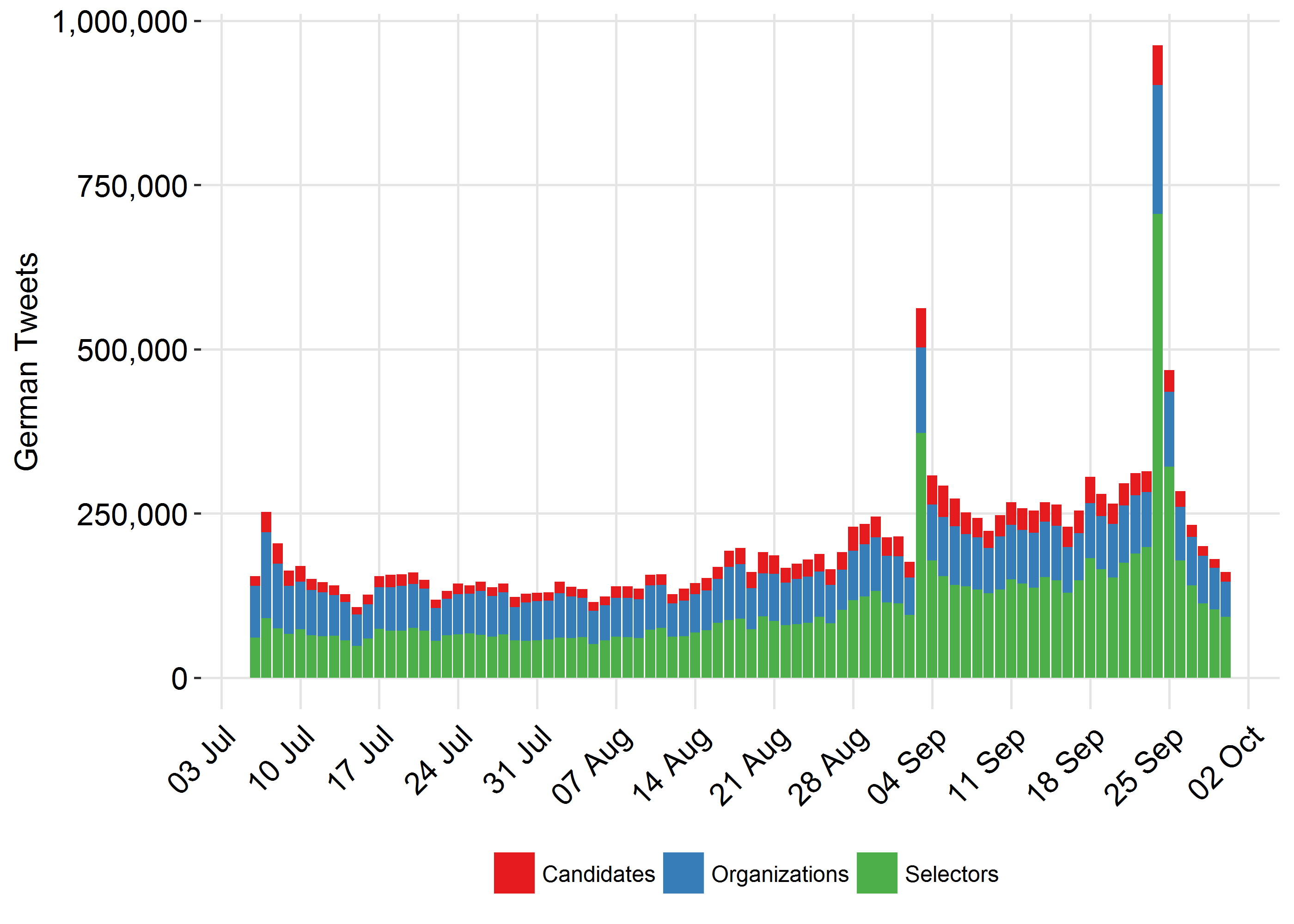}
        \caption{With German filter}
    \end{subfigure}
    \caption{Tweets in our dataset by selection logic.}
	\label{fig:tw_daily}
\end{figure}

Figure \ref{fig:tw_daily} displays the analogous data for Twitter, except that here we were also able to collect tweets containing keywords on political topics. Panel (a) shows the unfiltered dataset, that is all tweets we collected.\footnote{Note that this figure includes duplicates since a tweet will be collected twice if it addresses multiple target concepts. To give an example including a mention of a candidate and a topic selector: ``@MartinSchulz has won the \#tvduell''.} Just like in the case of Facebook, activity on Twitter is highest around election day. However, there are four more spikes in the time series: the first two, the G20 summit in Hamburg on 7/8 July 2017 and the terror attack in Barcelona on 17 August 2017, are not immediately related to the Bundestag campaign. Nonetheless, both events were related to political actors and topics that are prominently featured in our political communication space. The third noteworthy peak in Twitter activity is related to the ``TV Duell'' between the leading candidates Merkel (CDU) and Schulz (SPD) on 3 September 2017. 

The figures indicate that Twitter is the more event-driven medium, as the increases in volume related to ongoing political events are usually of a higher magnitude than on Facebook. It is an established finding in political communication that the Twitterverse intensively reacts to TV events during election campaigns. In fact, politicians themselves prefer Twitter over Facebook for ``dual screening'' purposes \citep{StierBleier.2017}.

As discussed before, all posts by media organizations were included in these figures based on the raw data. Many of the Facebook posts and tweets by media organizations are of course non-political and thus not immediately relevant to the campaign. We provide the account handles and IDs of these accounts \citep{DBK.2018}, which allows a researcher who has reconstructed our dataset to remove tweets, Facebook posts and interactions of these accounts if preferred for her analysis.

\subsection{Attention on Twitter}
This section explores the attention the Twitterverse devoted to different topics and actors. 
Twitter comes closest to an online ``public sphere'' that reacts to external events happening during a campaign. In contrast, Facebook is a medium dominated by reciprocal interactions, thus, the page owner has more influence on setting the agenda by choosing which topics to post on.

Figure \ref{fig:selectors} shows the most frequently mentioned political topics as predefined by the selectors found in Appendix \ref{appendix:selectors1_polity} to \ref{appendix:selectors4_parties}.\footnote{As we are interested in general attention patterns, we also include the appearance of selectors in combination with other words like ``\#noafd'' or when these are part of mentioned screennames like @AfDBerlin.} Most attention was devoted to the AfD, followed by mentions of the two larger parties (CDU, SPD, ``groko'' which stands for the governing ``grand coalition'' of the two) as well as their leading candidates Merkel and Schulz. We also find some generic terms like Bundestag or btw17, the foreign policy issue Turkey and the term refugees. Based on that figure, it seems that campaigning online resembles offline politics to a close extent -- most attention is devoted to the leading parties and candidates (see also \citealp{Yang.2017}), with the outlier being the AfD.

\begin{figure}
	\centering	
	\includegraphics[width=0.9\textwidth]{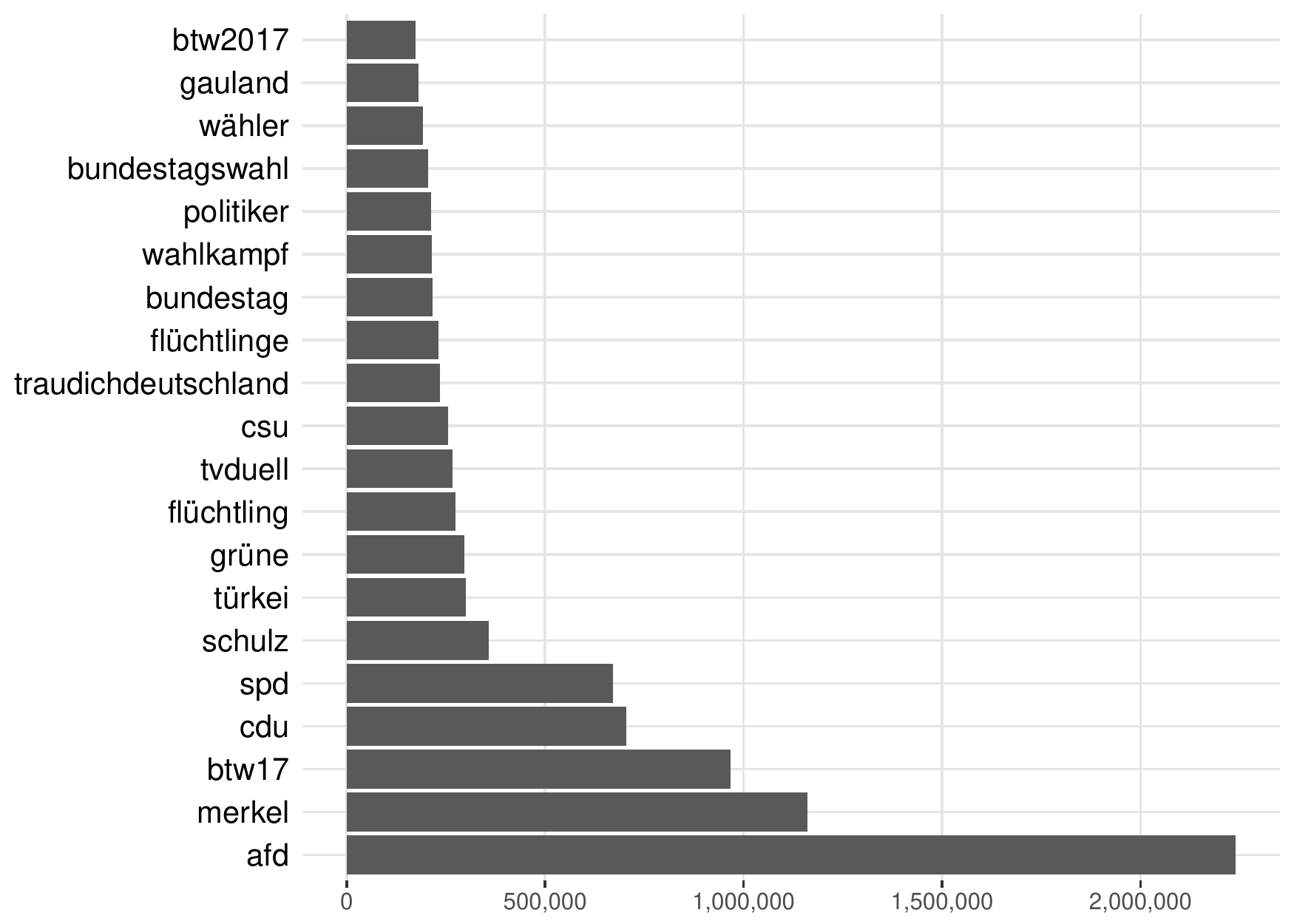}
	\caption{Top 20 mentioned selectors in our dataset.}
	\label{fig:selectors}
\end{figure}

\subsection{Activity and Resonance by Party}
Public and academic observers have put a particular focus on the use of social media by political parties and candidates. Thus, our report relies on our exhaustive data to investigate how political actors themselves have used social media and how their activities were received by citizens. 

In Table \ref{tab:table_accounts}, we first list the number of candidates who were running in the campaign for the seven main parties, next the number of candidates for which we found at least one Facebook or Twitter account and finally for how many candidates data could be retrieved during our research period (i.e., a candidate had a public and active account). 84\% of Bundestag candidates maintained a Facebook account, while 49\% had a presence on Twitter. Moreover, for 710 candidates, we found a second Facebook account for which we also collected data. There is a considerable difference in the retrieval rates for Facebook (48\%) and Twitter (67\%). The lower retrieval rates on Facebook are due to the fact that many accounts were set up as user profiles or private accounts (instead of accounts declared as Facebook pages). In contrast, almost all Twitter accounts were public, but approximately one third of candidates with a Twitter presence never tweeted during the campaign. 
\begin{table}
\centering
\caption{Social media adoption and number of retrieved accounts by party candidates}
\label{tab:table_accounts}
\begin{tabular}{lccccc}
Party & Candidates  & FB accounts & \makecell{public  \& \\ active FB} & TW accounts & \makecell{public  \& \\ active TW} \\ \hline
AfD       & 388  & 287    & 126                 & 139    & 75                  \\
CDU       & 477  & 400    & 196                 & 208    & 120                 \\
CSU       & 90   & 82     & 40                  & 55     & 33                  \\
FDP       & 367  & 329    & 160                 & 185    & 125                 \\
Grüne     & 360  & 313    & 114                 & 212    & 174                 \\
Linke     & 355  & 290    & 123                 & 163    & 109                 \\
SPD       & 479  & 408    & 252                 & 261    & 181                 \\ \hline
Total       & 2,516 & 2,109   & 1,011                & 1,222   & 817                
\end{tabular}
\end{table}

We compared these results to the ones reported by \citet{Schmidt.2017}. To be clear, the comparison is based on our final dataset that includes the candidate accounts we added from the Tagesspiegel and Open Knowledge Foundation databases. Schmidt and his team found a Facebook account (pages or profiles) for 1,849 candidates from the seven parties we focus on, compared to 2,109 in Table \ref{tab:table_accounts}. He reports 1,096 candidates with a Twitter account, our final dataset contains 1,222 Twitter accounts.

Next, we turn to the resonance that political actors attract on social media. This gives indications on whether the considerable staff and resources needed to maintain a social media presence are worth the effort. For this, we use measurements of social media influence that are established in the literature (e.g., \citep{Yang.2017,Kovic.2017,Nielsen.2013}. Note, however, that these metrics need to be interpreted with caution as they are particularly susceptible to automated activity by bots or astroturfing campaigns coordinated by humans \citep{Keller.2017}. It also has to be kept in mind that deletions before the 8-day rolling time window of our data collection are not included here. Deletions are, however, less frequent for party posts than for user generated content \citep{Bachl.2017}.

\begin{figure}
	\centering	
	\includegraphics[width=0.93\textwidth]{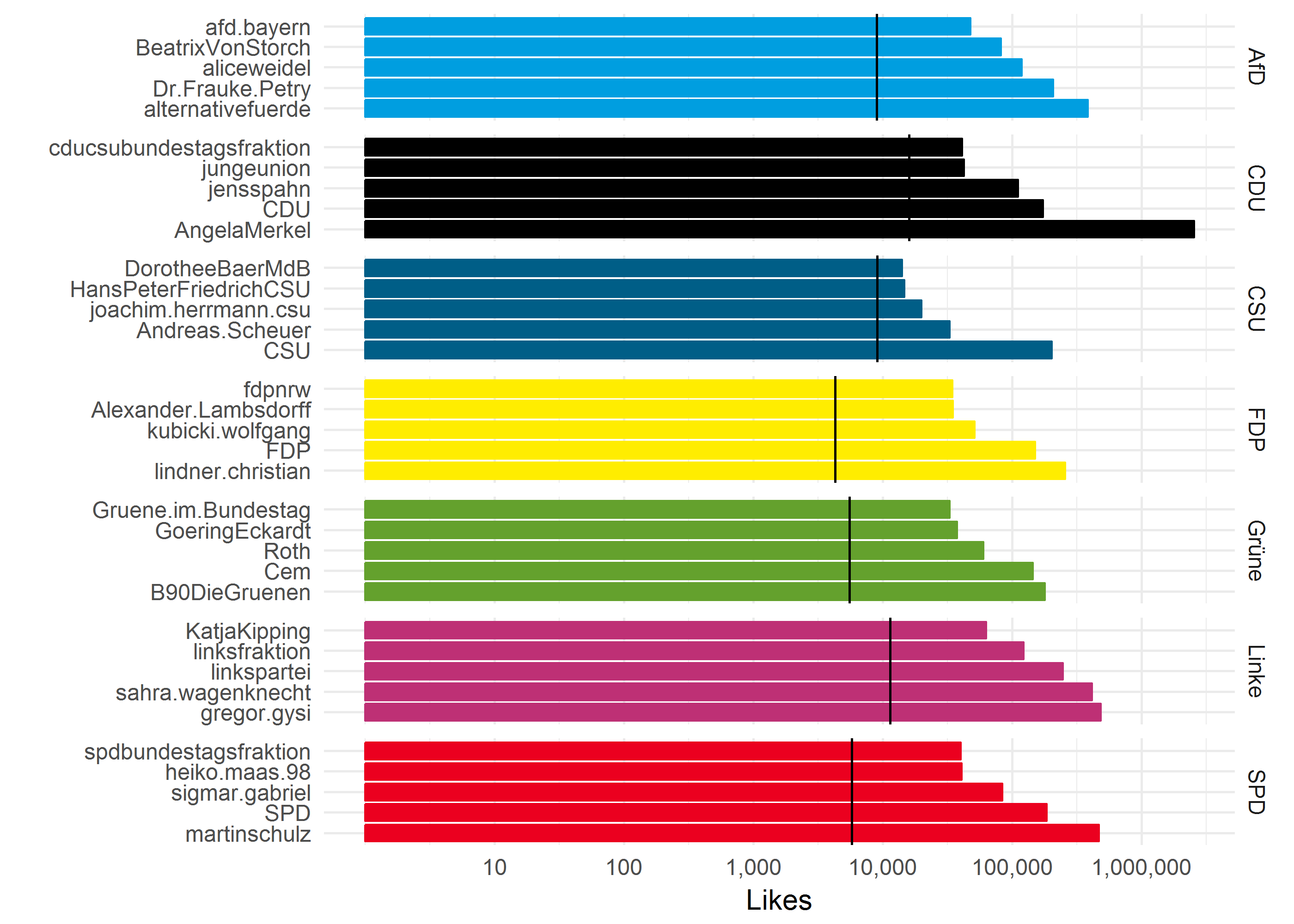}
	\caption{Top five Facebook accounts by page likes for each party on a logged scale. The black lines represent party averages.}
	\label{fig:partyaccs_fb}
\end{figure}

\begin{figure}
	\centering	
	\includegraphics[width=0.93\textwidth]{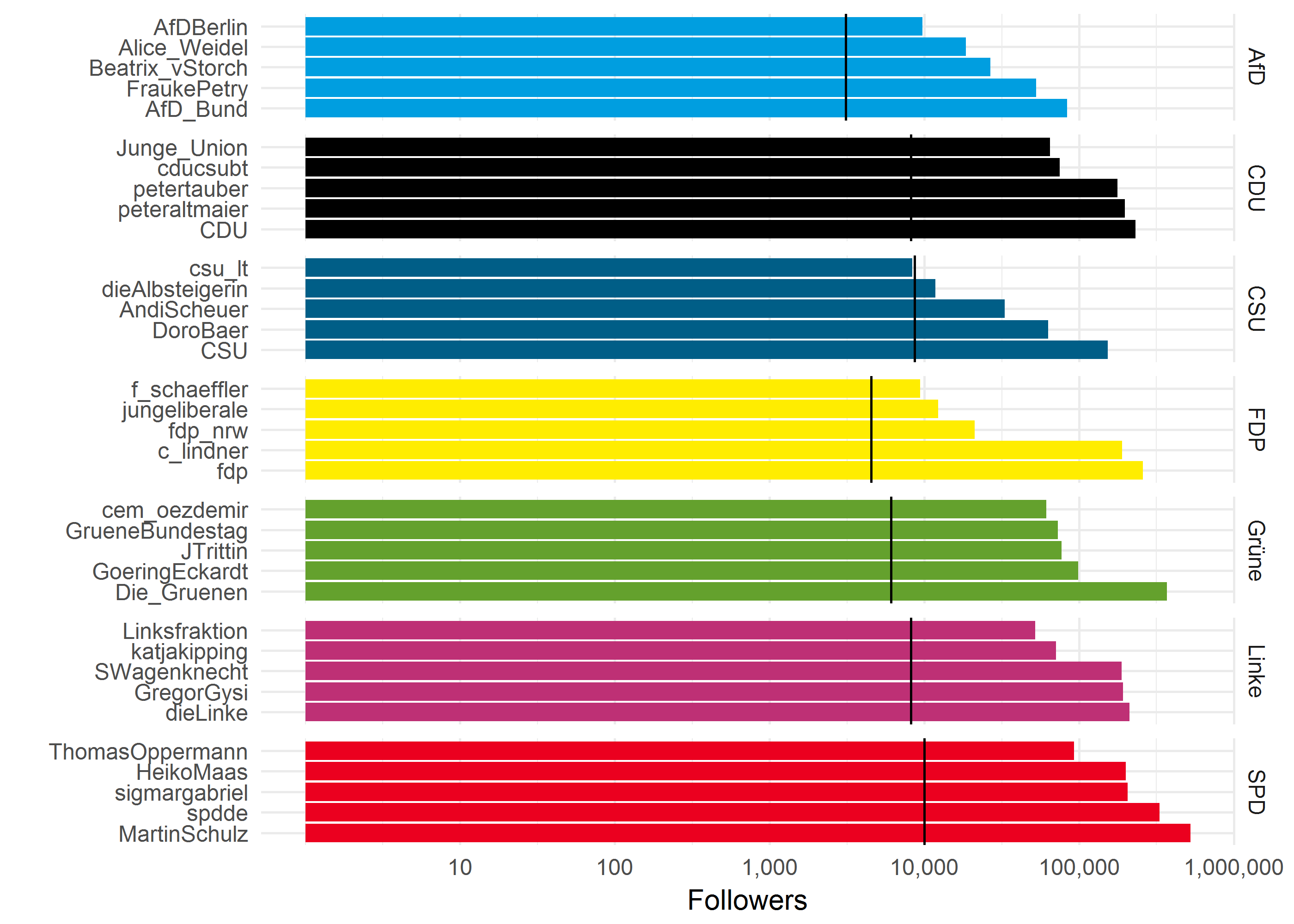}
	\caption{Top five Twitter accounts by the number of followers for each party on a logged scale. The black lines represent party averages.}
	\label{fig:partyaccs_tw}
\end{figure}

Figure \ref{fig:partyaccs_fb} shows the top five accounts per party according to how many Facebook page likes they have accumulated. 
Accounts of leading politicians and the main party accounts dominate the list. When looking at the averages per party, we cannot find clear-cut differences. However, this count highly depends on the number of included accounts and how many candidates are on party lists. Overall, there are no apparent asymmetries between parties in the number of page likes they attracted. 

In Figure \ref{fig:partyaccs_tw}, the same plot is displayed for the number of followers on Twitter, i.e., the people who regularly receive messages from political actors in their Twitter timelines. While Angela Merkel is the person with most page likes on Facebook, Martin Schulz leads the field in terms of Twitter followers. This is unsurprising, as Angela Merkel does not have a Twitter account and Schulz has gained international followers during his time as president of the European parliament. It is interesting to contrast the paltry follower numbers of leading and average AfD politicians to the prominent role of the AfD in the topics that were discussed on Twitter (see Figure \ref{fig:selectors}). Their lower number of followers notwithstanding, the party still overshadowed Twitter debates through various escalations in campaign rhetoric. This indicates that even though the AfD dominates Facebook and Twitter discourses, the tone of these debates (support for the AfD vs. negative depictions) might be rather different.

\begin{figure}[t]
	\centering	
	\includegraphics[width=0.94\textwidth]{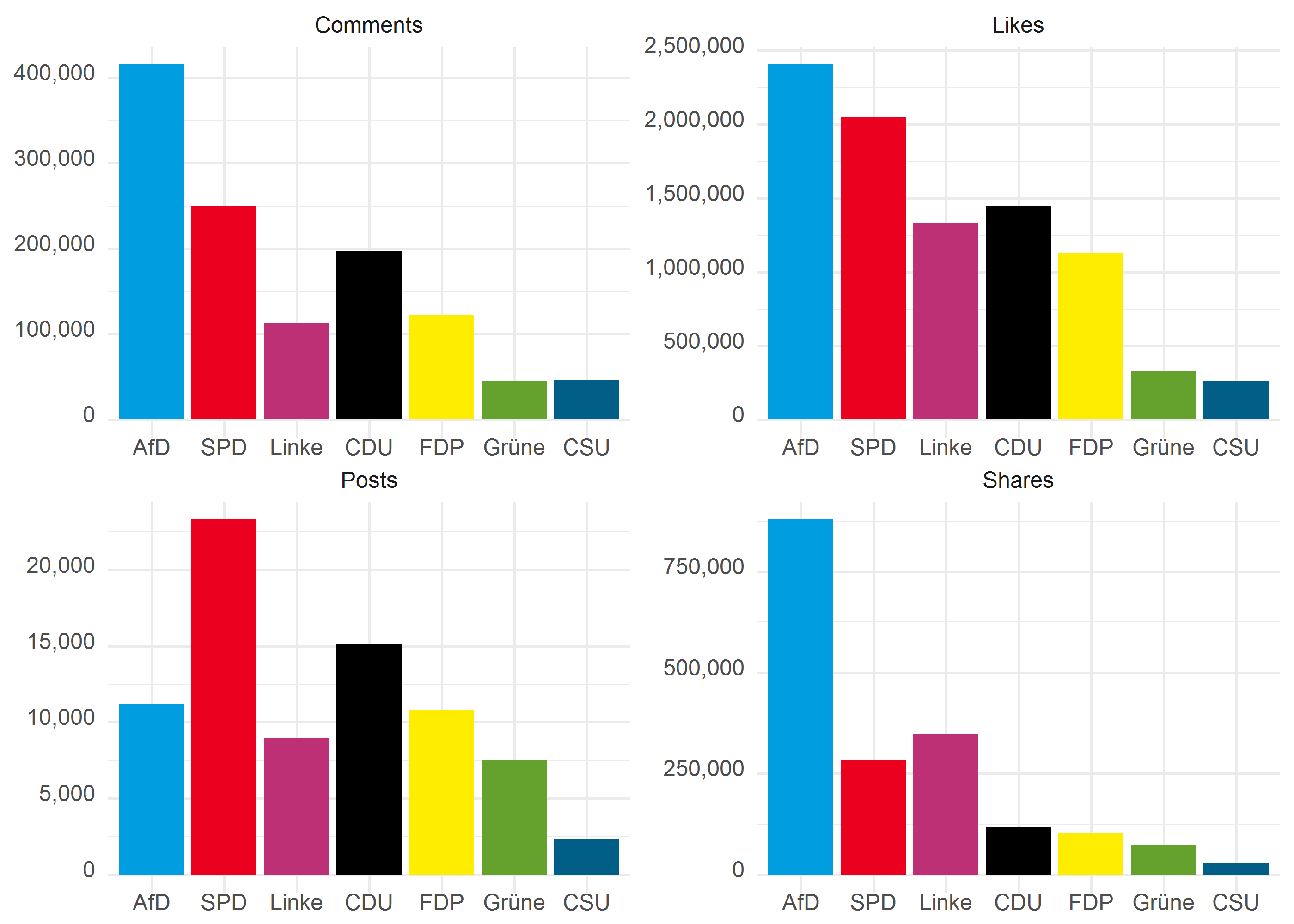}
	\caption{Aggregated Facebook posts and received engagement (comments, likes and shares) by party.}
	\label{fig:engagement_fb}
\end{figure}

In our final analysis, we move beyond the page level and investigate the actual engagement of audiences with contents produced by parties. This might be a better indicator for the ``success'' of online campaigns. As Figure \ref{fig:engagement_fb} makes strikingly clear, the AfD was by far the most successful party on Facebook in terms of engagement. Even though AfD candidates and party branches only posted an average number of posts, they received higher numbers of likes, comments and in particular shares than other parties. Likes and shares can be regarded as the clearest indications for political support, while comments tend to be distributed more evenly since they are also used to criticize account holders.

Overall, these results suggest that the AfD was successful in generating the most (positive or negative) attention during the election campaign. This echoes results from previous studies \citep{Stier.2017a,Kollanyi2017junk} and media reports on the AfD's extensive social media operations.\footnote{\url{http://www.sueddeutsche.de/politik/gezielte-grenzverletzungen-so-aggressiv-macht-die-afd-wahlkampf-auf-facebook-1.3664785-2}} Further research should investigate how AfD accounts are intertwined with supporter groups and so-called ``alternative media'' pages.\footnote{Account names of such online media can also be downloaded from \citet{DBK.2018}.} There is mounting evidence that especially on Facebook, a right-wing media ecosystem has emerged that continues to influence German online discourses after the election campaign.

\section{Data Sharing}
\label{sec:datasharing}

The data we collected is proprietary data owned by Facebook and Twitter. Because of that and due to privacy restrictions, it is not possible to share the raw data with other researchers or the public. However, there are possibilities to reconstruct social media datasets, which is important in order to promote reproducibility in digital research \citep{KinderKurlanda.2017}. Twitter allows the publication of tweet IDs for academic purposes. Using these unique numeric identifiers, researchers can reconstruct our Twitter dataset. We publish our masterlist of accounts which allows for the reconstruction of the Facebook dataset -- with certain limitations in ex post data availability \citep{Bachl.2017}. The necessary data for the reconstruction of the Twitter and Facebook datasets can be found at \citet{DBK.2018}.

\begin{figure}
	\centering	
	\includegraphics[width=1\textwidth]{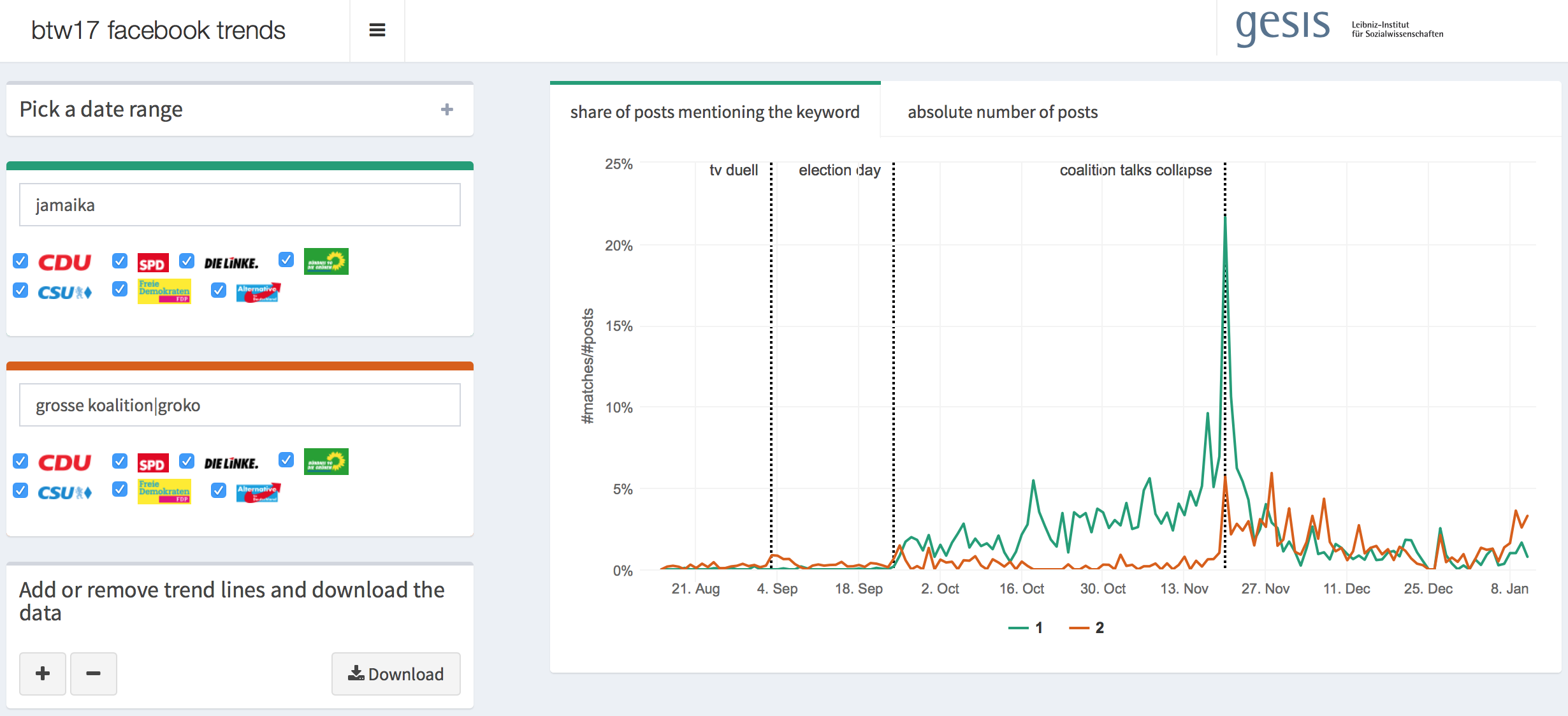}
	\caption{Interface of the online monitoring tool.}
	\label{fig:monitoring}
\end{figure}

While raw social media data cannot be shared, researchers can publish aggregated information from the metadata or text of social media posts. Thus, we provide an online access to our dataset that allows researchers to track the salience of topics of interest in customized searches. The resulting time series show how the salience of terms like ``merkel'' or ``afd'' develops over time on social media. In its current beta version, only posts by partisan actors, i.e., party branches and candidate accounts are included. In further versions, we will include data by regular users posting and commenting on political matters. 

Figure \ref{fig:monitoring} shows a screenshot of the monitoring platform with a time series from Facebook. We used keywords on the two most realistic coalition options ``jamaika'' and ``grosse koalition'' as search input in this example. The tool not only enables researchers to explore how social media reacts to unfolding political events, but also to download the data as relative frequencies over time. Moreover, the tool allows for a filtering of the data by party. The monitoring instrument can be accessed at \url{mediamonitoring.gesis.org}.

\section{Conclusions}
\label{sec:conclusion}

This project aimed to provide guidelines on how to collect social media data in a valid and reliable way, with a particular focus on the specifics of political communication. Conceptually, we first defined relevant target concepts prominent during an election campaign, namely candidates, organizations like party and media accounts and political topics. We then operationalized and collected data on these concepts in line with the possibilities and affordances of the two social media platforms Facebook and Twitter. Our temporal analysis showed that social media reacts considerably to campaign events like the TV debate or election night. In an analysis of party activities and their reception on social media, we found coherent evidence that the AfD was particularly successful in exploiting the potential of social media. Besides accounts of leading figures from other parties, the AfD generated the highest engagement on Facebook and was also the most talked about party on Twitter. As the political role of social media continues to grow, researchers will have to continue focusing on these partisan asymmetries that can only be identified if the data collection goes beyond a tiny collection of party accounts or campaign hashtags. It is of utmost importance to analyze political communication across all arenas of election campaigning, also in the long tail of candidates.

Based on these conceptual and infrastructural foundations, we now maintain a robust data collection scheme that will continue to monitor social media in Germany. Beyond political communication, the infrastructure can be activated for collecting data on various events and social phenomena ranging from election campaigns in other countries to non-political matters. The datasets on the Bundestag campaign can be reconstructed using the data we uploaded. The ongoing streaming of data can be searched and accessed via our monitoring tool. We think that these efforts represent a considerable step towards enhancing replicability and data sharing in social media research.

We also acknowledge several limitations and the need for further infrastructural investments to fully realize the value of digital behavioral data in the social sciences. The observational data we collected is limited in that it does not contain important contextual and individual-level information such as socio-demographic characteristics, voting behavior or political preferences. The lack of such data limits the theoretical and empirical value of digital behavioral data. Thus, we see a lot of potential in the linking of social media datasets with additional data sources like traditional surveys \citep{StierBleier.2017,Jungherr.2016bb} surveys of social media users \citep{Vaccari.2015}, candidate surveys \citep{Karlsen.2016,Quinlan.2017} or data from other sources such as real-time measurements during TV debates \citep{Maier.2016}. It is also clear that our previous data collections and the literature in general have mostly confined themselves to contentious political periods like election campaigns or offline protests. As a reaction to that, our data mining will continuously run throughout the current legislative session of the Bundestag. Having longitudinal data will allow social scientists to assess how every-day, less event-driven political processes manifest themselves on social media. Such data can serve as an analytical baseline and improve the understanding of whether established findings on online political communication still hold during less contentious political periods.

\bibliographystyle{apacite}
\bibliography{references}

\newpage
\setcounter{table}{0}
\renewcommand{\thetable}{A\arabic{table}}

\setcounter{figure}{0}
\renewcommand{\thefigure}{A\arabic{figure}}

\section{Appendix}

\begin{table}[ht]
\centering
\caption{Selectors: \textbf{Polity}}
\label{appendix:selectors1_polity}
\begin{tabular}{lll} \hline
bundestag             & bundesministerin & direktmandat             \\
länderfinanzausgleich & bundesminister   & wahlrecht                \\
bundeskanzler         & landtag          & landtagswahl             \\
kanzler               & bürgerentscheid  & verfassungsgericht       \\
bundeskanzlerin       & wahlsystem       & bundesverfassungsgericht \\
direktedemokratie     & bundesrat        & bverfg                   \\
direkte demokratie    & bundespräsident  & bundesregierung          \\
überhangmandat        & wähler           & verfassungsrichter       \\
bundestagswahl        &                  &             
\end{tabular}
\end{table}

\begin{table}[H]
\centering
\caption{Selectors: \textbf{Policy}}
\label{appendix:selectors2_policy}
\begin{tabular}{lll} \hline
finanzpolitik               & zensur                & energiewende            \\
steuerpolitik               & shadowbans            & klimawandel             \\
staatshaushalt              & shadow bans           & kulturpolitik           \\
wirtschaftslage             & atomkraft             & schulpolitik            \\
russland                    & atomenergie           & einkommensungleichheit  \\
türkei                      & antiatom              & steuergeschenk          \\
syrien                      & endlagerung           & wirtschaftspolitik      \\
arbeitslosigkeit            & gorleben              & integrationspolitik     \\
steuerreform                & atomausstieg          & rentenpolitik           \\
vorratsdatenspeicherung     & überwachung           & betreuungsgeld          \\
bankenkrise                 & datenschutz           & frauenquote             \\
finanzkrise                 & überwachung           & einkommensgerechtigkeit \\
ezb                         & islamist              & fachkräftemangel        \\
bankenaufsicht              & lohnpolitik           & gesundheitspolitik      \\
finanzmarktsteuer           & ehefueralle           & bildungspolitik         \\
agenda 2010                 & ehefüralle            & kinderarmut             \\
herdprämie                  & staatsdefizit         & arbeitsmarktpolitik     \\
einkommensschere            & ökosteuer             & altersarmut             \\
jugendarbeitslosigkeit      & hochschulpolitik      & solidaritätszuschlag    \\
forschungspolitik           & hartzIV               & auslandseinsatz         \\
verkehrspolitik             & hartz4                & elterngeld              \\
familienpolitik             & hartz 4               & freihandel              \\
bundeswehr                  & sozialpolitik         & flüchtling              \\
umweltpolitik               & soziale gerechtigkeit & fluechtling             \\
rentenreform                & mindestlohn           & flüchtlinge             \\
netzpolitik                 & wirtschaftskrise      & fluechtlinge            \\
Netzwerkdurchsetzungsgesetz & klimaschutz           & migranten               \\
netzdg                      & energiepolitik        &                        
\end{tabular}
\end{table}

\begin{table}
\centering
\caption{Selectors: \textbf{Politics / election campaign}}
\label{appendix:selectors3_politics}
\begin{tabular}{lll} \hline
wahl17         & wahljahr          & infratest             \\
btw2017        & umfrage                 & wahlbeteiligung       \\
btw17          & pegida                  & populismus            \\
wahlen2017     & forschungsgruppe wahlen & populistisch          \\
wahl2017       & lügenpresse             & politikverdrossenheit \\
kanzlerduell   & luegenpresse            & wahlversprechen       \\
tvduell        & forsa umfrage           & politiker             \\
elefantenrunde & politbarometer          & wahlwerbung           \\
parteiprogramm      & emnid                   & meinungsfreiheit      \\
schulzzug      & allensbach              & wahlkampf             \\
parteitag     &                 &     
\end{tabular}
\end{table}

\begin{table}
\centering
\caption{Selectors: \textbf{Parties \& leading politicians}}
\label{appendix:selectors4_parties}
\begin{tabular}{lll} \hline
grüne            & karrenbauer       & steinmeier           \\
bündnis90        & altmaier          & thorsten albig       \\
buendnis90       & bouffier          & woidke               \\
gruene           & groehe            & \#fdp                \\
goering-eckardt  & rainer haseloff   & lindner              \\
göring-eckardt   & stanislaw tillich & npd                  \\
goeringeckardt   & von der leyen     & piratenpartei        \\
göringeckardt    & gröhe             & koalition            \\
kretschmann      & linkspartei       & groko                \\
oezdemir         & bartsch           & grosse koalition     \\
özdemir          & wagenknecht       & große koalition      \\
hofreiter        & ramelow           & jamaikakoalition     \\
csu              & spd               & ampelkoalition       \\
seehofer         & schulz            & schwampel            \\
dobrindt         & hannelore kraft   & schwarzgrün          \\
christianschmidt & michaelmüller     & schwarzgelb          \\
afd              & heiko maas        & rotgrün              \\
frauke petry     & carsten sieling   & fedidwgugl           \\
gauland          & schwesig          & traudichdeutschland  \\
weidel           & sigmar gabriel    & holdirdeinlandzurück \\
cdu              & barbara hendricks & denkenwirneu         \\
merkel           & sellering         & lustauflinks         \\
schäuble         & olaf scholz       & darumgrün            \\
schaeuble        & malu dreyer       & darumgruen           \\
maiziere         & nahles            & zeitfürmartin        \\
johanna wanka    & stephan weil      & zeitfuermartin      
\end{tabular}
\end{table}


\end{document}